\documentclass[3p,times,twocolumn]{elsarticle}

\usepackage{ecrc}

\newcommand{\sqrts}{\sqrt{s}}
\newcommand{\sqrtsNN}{\sqrt{s_{\scriptscriptstyle \rm NN}}}

\newcommand{\gev}{\mathrm{GeV}}

\newcommand{\tev}{\mathrm{TeV}}

\newcommand{\PbPb}{\mbox{Pb--Pb}}
\newcommand{\pPb}{\mbox{p--Pb}}

\newcommand{\pt}{p_{\rm T}}

\newcommand{\Dzero}{{\rm D^0}}

\newcommand{\Ds}{{\rm D_{s}}}
\newcommand{\Lc}{\Lambda_{\rm c}}

\newcommand{\Jpsi}{{\rm J}/\psi}
\newcommand{\ccbar}{{\rm c}\bar{\rm c}}

\renewcommand{\PbPb}{\mbox{Pb--Pb}}
\newcommand{\Npart}{N_{\rm part}}
\newcommand{\Ncoll}{N_{\rm coll}}

\newcommand{\RpPb}{R_{\rm pPb}}
\newcommand{\RAA}{R_{\rm AA}}

\newcommand{\vtwo}{v_{2}}

\volume{00}

\firstpage{1}

\journalname{Nuclear Physics B Proceedings Supplement}

\runauth{A. Rossi}

\jid{nppp}

\jnltitlelogo{Nuclear Physics B Proceedings Supplement}

\usepackage{amssymb}

\usepackage[figuresright]{rotating}

\begin{document}

\begin{frontmatter}



\dochead{}

\title{Experiment summary}


\author[unipd]{A. Rossi}

\address[unipd]{Padua University and INFN}

\begin{abstract}
The measurement of the production of particles coming from hard scattering processes covers a fundamental role in the 
characterization of the system formed in heavy-ion collisions, allowing to probe the microscopic processes underlying 
the interaction of high energy partons with the medium. An impressive amount of measurements related to jet, quarkonia, open heavy 
flavor, and electroweak signal production in nucleus-nucleus as well as p(d)-nucleus collisions was delivered by experiments 
at RHIC and LHC in past years. In these proceedings, the main experimental results presented during the Hard Probes conference are summarized.
\end{abstract}


\end{frontmatter}


During the last two decades, high-energy nuclear physics has seen a tremendous progress. 
Several evidences were collected supporting the presence of a phase transition
to a medium of deconfined quarks and gluons (Quark-Gluon Plasma, QGP) in relativistic heavy-ion collisions. With ``hard-probes'', i.e. with particles produced
in hard-scattering processes with large momentum transfer, we can (with a maybe improper expression) resolve the medium constituents
and connect global medium properties, determining the evolution of the medium as an extended system, to the
parameters (e.g. transport coefficients) characterizing ``local'' partonic interactions. Thus, hard-probes represent
a unique opportunity to achieve a microscopic description of the medium. In these proceedings, the experimental results 
presented during the $7^{\rm th}$ edition of the Hard Probes conference are 
reviewed. This summary is not exhaustive: in particular, for a review of results related to dilepton production 
and direct photons at low $\pt$, see~\cite{Geurts,Novitzky}.
\section{Electromagnetic probes for validating Glauber-model scaling}
\label{sec:EM}
The measurement of the nuclear modification factors ($\RAA$) of W and Z bosons and of isolated photons at high transverse momenta
represents a fundamental and unique opportunity to measure initial state effects directly in nucleus-nucleus
collisions and to support the assumption that, in the absence of nuclear effects, the production of signals produced in    
hard scatterings of partons from the colliding nuclei scales with the number of binary nucleon-nucleon collisions estimated
with the Glauber model. The ATLAS experiment measured the production of isolated photons as a function 
of transverse momentum in the range $22<\pt<280~\gev/c$ in $\PbPb$ collisions at $\sqrtsNN=2.76~\tev$ in four centrality intervals, from 0 to 80\%,
and in two pseudorapidity intervals ($|\eta|<1.37$ and $1.37<\eta<2.37$)~\cite{ATLASphotonsPbPb}. The $\pt$-differential spectra, divided by 
the average nuclear thickness function, 
are described within uncertainties by NLO pQCD calculations implemented in the program 
JETPHOX 1.3~\cite{JETPHOX}, with and without incorporating
EPS09 nuclear modification of nucleon PDF~\cite{EPS09}. The pseudorapidity distribution of positively and negatively charged 
electrons and muons from W decay and the centrality dependence of their production are well reproduced by expectations 
based on a realistic cocktail of POWHEG~\cite{POWHEGatlasW} simulations of W production in proton-proton, proton-neutron, and neutron-neutron 
collisions, as observed by ATLAS~\cite{ATLASbosonWPbPb}. While the usage of proper proton and neutron PDFs is 
fundamental to account for the different valence quark composition of proton and neutrons and describe the data, there is no evidence, considering
the uncertainty of both data points and theoretical predictions, for a better description of the data if EPS09 parametrization 
of nuclear PDF is used. Agreement with binary scaling was also found for 
Z-boson production, measured by CMS in the dimuon ($|y|<2.0$) and dielectron ($|y|<1.44$) channels in $0<\pt<100~\gev/c$, as a function of transverse momentum 
in minimum-bias $\PbPb$ collisions and, $\pt$-integrated, as a function of rapidity and centrality~\cite{CMSbosonZPbPb}. In general, although
the task of validating binary scaling seems at certain extent completed, more precision
is needed on all measurements of high-energy electroweak signals to further constrain modes including initial-state effects.
\section{Results related to the investigation of partonic in-medium energy loss}
High-energy partons lose energy interacting with the medium constituents via both radiative (gluon emission) and collisional
processes. The characterization of the partonic energy loss, from which information can be obtained
on the transport coefficients of the medium, is a major goal to achieve a microscopic description 
of the medium. The study of the modification of production rates and internal structure of jets over a wide kinematic 
range is a key step to quantify the amount and the spatial distribution of the energy lost. A different (smaller) energy loss 
is predicted for quarks than gluons and for heavy-quarks than light quarks. Stringent constraints to models can be set
by comparing the modification to the production rates of heavy-flavour and light-flavour particles.
\subsection{Jet suppression and jet structure}
The nuclear modification factor of inclusive jets measured by ATLAS~\cite{ATLASjetPbPb} indicates a suppression by a factor about two of jet ($R=0.4$) production
from $\pt=50~\gev/c$ up to $400~\gev/c$ in the 0-10\% most central $\PbPb$ collisions, with none or little dependence on $\pt$. The suppression, which
does not show a significant difference in the rapidity ranges $0.3<|y|<0.8$ and $1.2<|y|<2.1$, reduces 
with centrality and $\RAA$ is about 0.8 in the 60-80\% centrality range. In the 0-10\% most central collisions, ALICE observed a suppression by a 
factor about 3 down to $\pt=40~\gev/c$ for jets with $|y|<0.5$ reconstructed with $R=0.2$ and requiring a charged particle with $\pt>5~\gev/c$
among the jet constituents~\cite{ALICEjetsPbPb,ALICEjetsMegan}. The two measurements imply a strong interaction of partons with the medium up to very high momentum and indicate that 
at least part of the energy lost is dissipated outside the jet cone. Within uncertainties the measurements are reproduced by models. 

The nuclear modification factor of charged particles in the 0-5\% most central $\PbPb$ collisions was measured by ATLAS up to $\pt=200~\gev/c$ ($\RAA\sim0.65$)
with sufficient precision to appreciate an inflection point around $\pt=50~\gev/c$ ($\RAA\sim0.5$) after which $\RAA$ rises with $\pt$ is 
milder~\cite{ATLASchargedPartPbPb}. Compatible values were observed by ALICE, ATLAS, and CMS in the overlapping range of their
measurements, all showing a minimum of $\RAA\sim0.15$ at $\pt\sim6~\gev/c$~\cite{ALICEchargedPartPbPb,ATLASchargedPartPbPb,CMSchargedPartPbPb}. Charged 
particles with high momentum are in most cases leading hadrons of jets. Therefore, the measured $RAA$
indicates a suppression of the higher $\pt$ part of jet constituents that composes the jet ``core'', with particle momenta closer to the jet axis 
and carrying a large fraction of jet energy. ALICE has found that the relative abundances of pions and protons for $\pt>10~\gev/c$, and pions and 
kaons for $\pt>4~\gev/c$ are the same within uncertainties in pp collisions
and in central $\PbPb$ collisions~\cite{ALICEidPartPbPb}. This implies that the hadrochemical composition 
of the jet ``core'' is unaltered despite the strong suppression
of jet production. ALICE measured the $\Lambda/K^{0}_{\rm s}$ ratio in jets and found it compatible within (large) uncertainties with the 
values measured in pp and $\pPb$ collisions and significantly smaller than the inclusive value measured in central 
$\PbPb$ collisions~\cite{ALICELambdaOverKaonVit}, suggesting
that radial flow and hadron formation via coalescence do not affect the kinematic properties and particle composition 
of jet constituents. However, more precise measurements exploring a wider kinematic range in terms of jet 
momenta are needed for concluding. 

The measurement of the jet momentum fraction ($z$) carried by charged particles~\cite{CMSfragFunctionPbPb,ATLASfragmFunctPbPb} 
and the analysis of jet shape modification~\cite{CMSjetShapePbPb,ATLASjetShapePbPb} performed by ATLAS and CMS in recent years
suggest a small modification of jet anatomy, consisting of an enhancement of the multiplicity of low-$\pt$ constituents with small $z$ and 
at large angles with respect to the jet cone. The jet core, composed of particles at high $z$, contributing up to 85\% of jet energy,
and contained inside a cone of $R=0.1$ with respect to the jet axis, does not show significant modifications in central $\PbPb$ collisions
with respect to pp collisions. Further insight into the spatial distribution and kinematic properties of the radiated energy is provided by dijet analyses. Since
the energy loss increases with the distance covered by the parton in the medium, particles and jets with high-$\pt$
typically probe partons produced in the external layers of the fireball and going outward. Therefore, the study
of jets recoiling with respect to a high energy signal (jet or single particle) allows to probe the energy loss of partons covering 
long distances in the medium and suffering a stronger loss of energy. Both ATLAS and CMS found a significantly larger 
imbalance of the transverse momenta of the leading and sub-leading jets produced in central $\PbPb$ collisions
with respect to pp collisions~\cite{ATLASdijetMomIbalancePbPb,CMSdijetMomImabPbPbFirst}. In order to better qualify
the observed imbalance, CMS selected events with large $\pt$-asymmetry $A_{J}=(\pt^{\rm Lead}-\pt^{\rm SubLead})/(\pt^{\rm Lead}+\pt^{\rm SubLead})$
and, by analyzing the angular distribution of tracks with respect to the dijet axis for several $\pt$ intervals, observed that
the additional imbalance arises from a suppression of the high-$\pt$ constituents of 
the sub-leading jet, close to the dijet axis~\cite{CMSmissingPt,CMSyenjielee}. The 
``missing energy'' is recovered by particles at low $\pt$ (smaller than $2~\gev/c$) and at large angles. 
Additional evidence in this direction is provided by the analysis
of angular correlations between jets and charged particles~\cite{CMSyenjielee}. The yield of particles with $\pt<3(2)~\gev/c$ associated 
to the sub-leading (leading) jet
is significantly larger in central $\PbPb$ collisions than in pp collisions and the width of the $\Delta\eta$ and $\Delta\varphi$ distributions
is wider in the nuclear case. No significant difference can be appreciated for particles with higher $\pt$. The trend
is qualitatively consistent with the picture emerging from the study of the missing energy distribution and, for leading jets, with 
the measurement of the fragmentation functions mentioned above.  
At RHIC energies, STAR explored a new technique for reducing the combinatorial background for the identification of dijet 
pairs, that allows to reconstruct jets with a loose selection
on the minimum track momentum as well as with a larger radius~\cite{STARdijetJetFinder}. %
The $A_{J}$ distribution measured with $R=0.2$ 
and $\pt^{\rm cut}=2~\gev/c$ in Au--Au collisions for $\pt^{\rm Lead}>20~\gev/c$ and $\pt^{\rm SubLead}>10~\gev/c$ shows a stronger 
imbalance with respect to the distribution measured in pp collisions. Conversely, with $R=0.4$
and $\pt^{\rm cut}=0.2~\gev/c$ $A_{J}$ distributions compatible within uncertainties are observed in the two colliding systems. This trend
may suggest that, contrary to what observed at LHC, at RHIC energy the energy lost is dissipated inside a cone with $R=0.4$, thus relatively
close to the jet axis. However, more differential analyses in terms of $\pt^{\rm Lead}$ and $\pt^{\rm SubLead}$ are needed 
to better understand the bias induced by the jet kinematic selection, and obtain a deeper interpretation of the 
results. 

The modulation of $A_{J}$ with the angle between the leading jet and the event plane was studied at LHC by CMS
as a function of collision centrality. The coefficient of the second order Fourier harmonic, averaged
over the six centrality intervals considered in the range 0-80\%, shows 
a slightly negative value ($-0.010 \pm 0.004$), which suggests a stronger imbalance for jet pairs
oriented ``out-of-plane'', in qualitative agreement with the expectation of a larger energy loss from
the longer path covered by the recoiling parton in the medium. 

Novel and promising results for investigating the suppression and properties of recoiling jets down to low jet $\pt$ and 
without introducing fragmentation biases were obtained
by ALICE and STAR with the analysis of azimuthal correlations of high-$\pt$ particles and jets~\cite{ALICEjetsMegan,ALICEhJet,STARhJet}. ALICE removes
the contribution of combinatorial background jets by subtracting the spectrum of jets recoiling with respect 
to a lower-$\pt$ hadron ($8<\pt<9~\gev/c$) and compares the resulting spectrum to expectations from PYTHIA. STAR, instead, used 
uncorrelated hadron-jet pairs obtained with event mixing and compare the background-subtracted spectrum obtained in central collisions
with that in peripheral collisions. Both experiments observed a significant ``additional'' suppression of the recoiling jets, and though
the much steeper initial jet $\pt$-spectrum shapes at RHIC energies than at LHC, and the different kinematic selections 
prevent the possibility of a direct comparison, a constant energy loss of about $8~\gev/c$ in the $\pt$ ranges considered is suggested
by both measurements. By comparing the recoiling spectra obtained with $R=0.2$ and $R=0.5$, ALICE did not observe 
any evidence within uncertainties for a broadening of the jet structure with respect to what expected from PYTHIA. 
The width of the away-side peak of the $\Delta\varphi$ distribution in ALICE data and in PYTHIA are compatible, constraining the size of possible medium
induced acoplanarity. The rate of single large-angle (Moli\`ere-like) scattering was estimated from the analysis of the yield of 
associated jets to be not significant, though still with a limited precision.
\subsection{Heavy-flavour quarks: the advantage of identity}
A suppression of b-jet production increasing with centrality up to more than a factor 2 in the 10\% most central $\PbPb$ collisions 
was measured by CMS for $\pt>80~\gev/c$~\cite{CMSbjetRAA}. The b-jet $\RAA$ is compatible within uncertainties with 
that of inclusive jets and described by the model by J.~Huang~et~al.~\cite{bjetsVitev} that contains a dependence 
of the energy loss from the quark mass but predicts a possible higher $\RAA$ for b-jets with $\pt$ below 
$\sim~60~\gev/c$. An indication of a mass dependence of the energy lass derives from the comparison~\cite{ALICEcristina} of the centrality
dependence of $\RAA$ of prompt D-mesons in $8<\pt<16~\gev/c$ measured by ALICE~\cite{ALICEDraavsNpart}, and that of non-prompt 
$\Jpsi$ measured by CMS in $6.5<\pt<30~\gev/c$~\cite{CMSjpsifromBRAA}. The $\pt$ range of the D meson measurement
was defined in order to have a significant overlap and a similar median between the momentum distribution of D mesons and of B mesons decaying to $\Jpsi$ with
transverse momentum in the range of CMS measurement. Therefore, the $3.5\sigma$ difference between the average of the $\RAA$ values in 0-10\% and 10-20\% 
provides an indication for $\RAA({\rm B})>\RAA({\rm D})$. The values and centrality trend of the nuclear modification factors
are described within uncertainties by models (e.g.~\cite{DjordjevicELoss}) in which the $\RAA$ difference derives mostly
from the dependence of energy loss on the Casimir factor and on the quark mass rather than from the different charm and beauty quark 
pp $\pt$-differential spectra and fragmentation. The result can therefore be interpreted
as an indication for a smaller energy loss for the more massive beauty quark than for charm. CMS presented
a first observation of a ${\rm B}^{+}\rightarrow\Jpsi{\rm K^{+}}$ signal~\cite{CMSBpbpb}: the increase statistics expected
from run 2 at the LHC should allow for a first measurement of B-meson $\RAA$, that will be extremely important to 
access the b-quark kinematics more closely. 

The D-meson $\RAA$ is compatible with that of charged pion in a wide
momentum range, from 1 to $36~\gev/c$ (assuming the same $\RAA$ for charged pions and charged particles for $\pt>16~\gev/c$), as well
as at high $\pt$ as a function of centrality~\cite{ALICEcristina}. Models predicting a larger energy loss
for gluons than quarks and a slightly smaller value for the not-so-heavy charm quark obtain similar $\RAA$ values (e.g.~\cite{DjordjevicELoss})
due to the different charm, light-quark and gluon pp $\pt$-spectra and fragmentation functions. Stringent constraints on 
models describing charm quark interaction with the medium are set by the comparing data and predictions simultaneously for 
$\RAA$ and elliptic flow ($\vtwo$)~\cite{ALICEDmesonv2longpaper}. The latter is better described, at low $\pt$ by models including a realistic medium evolution and 
mechanisms (e.g. collisional energy loss and hadronization via coalescence) that transfer to charm quarks the elliptic flow induced during the 
system expansion. Deep insight into the relevance of coalescence for charm hadronization can be obtained by measuring $\Ds$ and
$\Lc$ $\RAA$ and $\vtwo$. ALICE observed an intriguing trend of ${\rm D_{s}}~\RAA$ for $\pt<8~\gev/c$: more precise measurements are needed
to clarify whether ${\rm D_{s}}$ production is enhanced with respect to non-strange D mesons in heavy-ion collisions, which would constitute a clear evidence
of coalescence of charm quark with strange quarks in the medium. In central Au--Au collisions at $\sqrtsNN=200~\gev$ at 
RHIC, STAR measured the $\Dzero$ nuclear modification factor as a function of $\pt$ down to $\pt=0$, observing a 
tendency for $\RAA>1$ in $1<\pt<2~\gev/c$~\cite{STARHF,STARDprl}, suggesting a different trend with respect to that measured by ALICE in Pb--Pb collisions
at $\sqrtsNN=2.76~\tev$. For a proper comparison, the possible stronger shadowing and the less steep spectrum at LHC, as
well as the potentially different impact of radial flow and hadronization via coalescence should be taken into account. The TAMU model~\cite{TAMU}
can reproduce, within uncertainties the two results. Therefore, more than representing a possible inconsistency, the difference
between LHC and RHIC data highlight a unique opportunity to exploit as a lever arm the distinct characteristics of charm production
at the two energies and get insight into several physics processes at play at both energies. With the newly installed
Heavy-Flavour Tracker detector at STAR and the next runs at the LHC (especially after ALICE detector upgrade in 2019)  
precise $\pt$-differential measurements of non-strange D mesons, $\Ds$, and $\Lc$ $\RAA$ and $\vtwo$ will become possible at both RHIC
and LHC energies. At LHC, the LHCb Collaboration decided on a significant extension of its heavy-ion physics
programme, foreseeing a possible participation to $\PbPb$ data taking, as well as to serve as a fix target experiment for studying
collisions of Pb (proton) beams with gas of various atomic species at $\sqrtsNN\sim 70(115)~\gev/c$~\cite{LHCbHIC}.
\section{Quarkonia: Debye screening and recombination}
The observation of $\Jpsi$ suppression at the SPS~\cite{JpsiSPS}, predicted as an effect arising from melting of charmonia states due to the Debye-like screening
of $\ccbar$ attractive potential in the presence of a colour medium~\cite{MatsuiSatz}, constituted a milestone for the discovery of the QGP. After 
the puzzling observation of a similar suppression at the higher energy density of Au--Au collisions at RHIC~\cite{JpsiPHENIX2007}, the observation of 
a smaller suppression of $\Jpsi$ production 
in central $\PbPb$ collisions at LHC compared to what measured at RHIC~\cite{JpsiALICE2012}, pointing to an additional formation 
of $\Jpsi$ from the combination of charm and anti-charm quarks originated in different hard-scattering processes, constitutes 
a second milestone provided by charmonia for heavy-ion physics. ALICE presented further studies for characterizing $\Jpsi$ suppression
trend as a function of the collision centrality for $\Jpsi$ transverse momenta in the ranges $\pt<2~\gev/c$, $2<\pt<5~\gev/c$, and
$5<\pt<8~\gev/c$~\cite{ALICEjpsi2015,ALICElardeux}. The results support a scenario with formation of $\Jpsi$ from recombination at low $\pt$
and  suppression due to Debye screening as the main effect at high $\pt$, where $\RAA$ decreases with centrality. A somehow unexpected
outcome of this study is the observation of an excess of $\Jpsi$ production with $\pt<0.3~\gev/c$ in peripheral collisions, with
a $\pt$ spectrum resembling that of $\Jpsi$ photo-production, studied in ultra-peripheral collisions with impact parameter larger
than twice the Pb-nucleus radius.   

CMS presented an update of the nuclear modification factors of bottomonia states, using a new reference 
from pp collisions at $\sqrts=2.76~\tev$ and a larger $\PbPb$ data sample. The ``sequential melting'' already
observed in the previous measurement~\cite{CMSbottomonia2012} is confirmed, but the new measurement indicates that 
$\Upsilon({\rm 1S})$ $\RAA$ decreases with centrality down to the significantly low value of 0.3 in central 
collisions. Considering that the feed-down contribution from excited $\Upsilon$ states and $\chi_{\rm b}$ is around
30\%, this strong suppression may point to a melting of the ground state itself~\cite{camelia}. In minimum-bias 
collisions, both $\Upsilon({\rm 1S})$ and $\Upsilon({\rm 2S})$ $\RAA$ do not show a significant 
dependence on $\pt$ in the range $0<\pt<20~\gev/c$ and, considering also ALICE data, on rapidity in the range $0<|y|<4$, setting 
stringent constraints for a statistical regeneration
of $\Upsilon({\rm 1S})$ that should be more significant at central rapidity.

At RHIC, data from U--U collisions at $\sqrtsNN=193~\gev$ allowed to extend 
charmonia and bottomonia measurements to higher $\Npart$ values, in a range in which both an increase Debye screening 
(due to the higher energy density) as well as recombination (due to the higher heavy-quark pairs multiplicity from the larger 
$\Ncoll$ value) could be expected. PHENIX measured compatible values and centrality trend of $\Jpsi$ $\RAA$ in U--U collisions and in 
Au--Au, within uncertainties~\cite{IordanovaPHENIX}. STAR measured compatible results for the centrality-integrated $\RAA$ of $\Upsilon({\rm 1S+2S+3S})$
and for the ground state $\Upsilon({\rm 1S})$ alone in the two collision systems~\cite{STARupsilonAuAu}. A similar centrality trend was also observed, with
$\Upsilon({\rm 1S})$ $\RAA$ reaching about 0.4 in central U--U collisions, yielding a first indication for a suppression of $\Upsilon({\rm 1S})$ production
at RHIC energies.
\section{Results from small collision systems: measuring cold nuclear matter effects and beyond}
The analysis of particle production in small systems, like those provided by p--Pb and d--Au collisions, is
fundamental for measuring initial and final state ``cold'' nuclear matter effects
and achieve a correct interpretation of what observed in nucleus--nucleus collisions. The main effect expected to 
modify the production of particles coming from hard-scattering processes at LHC energies
is nuclear shadowing, i.e. the suppression of gluon parton distribution function (PDF) at low Bjorken-$x$ values
in the nucleus with respect to the proton. Compatibility within uncertainties of $\RpPb$ with unity, as well as with
predictions including the EPS09 parametrization of nuclear PDF~\cite{EPS09}, was reported for many observables, e.g.
inclusive jets~\cite{ATLASjetRpPb}, b-jets, B mesons~\cite{CMSBpbpb}, %
D mesons~\cite{ALICEcristina,ALICEDmesonPPb}. The $\pt$- and y-differential cross sections
of non-prompt $\Jpsi$ with $8<\pt<30~\gev/c$ are in agreement within uncertainties with expectations from FONLL; the
forward-to-backward ratio measured is compatible with unity for $\pt>8~\gev/c$~\cite{CMSBpbpb,ATLASjpsipPb}. Compatible results 
were found also between the measurement of azimuthal correlation of D mesons and charged particles produced in the event
in pp collisions at $\sqrts=7~\tev$ and $\pPb$ collisions at $\sqrtsNN=5.02~\tev$, after 
the subtraction of a baseline representing the physical minimum of the distributions~\cite{Jitendra}. 
The dijet $\langle k_{{\rm T}y}\rangle$ distribution measured by ALICE as a function of the ``trigger'' 
jet $\pt$ is reproduced by simulations of pp collisions done with PYTHIA8~\cite{ALICEjetsMegan,ALICEjetktypPb}. 

The observation of charged-particle $\RpPb>1$ for $\pt\gtrsim30~\gev/c$ by CMS~\cite{CMSRpPb} and ATLAS~\cite{ATLASjetRpPb}, with
a trend rising with $\pt$ and contrasting, though compatible within uncertainty, with $\RpPb\sim 1$ suggested by ALICE 
measurement~\cite{ALICERpPb}, represented one of the unsolved puzzled of the last two years. An effort was made by the experiments
in order to understand the origin of the difference among the results and to measure the distribution of 
jet momentum fraction ($z$) carried by charged particles, expected to be modified in order to reconcile the charged 
particle $\RpPb$ with jet $\RpPb\sim1$. CMS showed the ratio between a new reference pp cross-section with respect to that used for the 
published $\RpPb$~\cite{CMSnewRpPb}: the difference could bring the $\RpPb$ of charged particle significantly closer 
to unity. The ratio of the $z$ distribution in the range $0<-\log(z)<5$ measured in $\PbPb$ and in pp collisions was found compatible with unity
by CMS independently from jet $\pt$ in the range $60<\pt^{\rm jet}<200~\gev/c$, while ATLAS observed a hardening 
of the $z$ distribution for $80<\pt^{\rm jet}<250~\gev/c$~\cite{ATLASjetRpPb}. The discrepancy between the two results derives likely
from the different methodology used to define the reference and it is being investigating.

Further insight into initial state effects is obtained by studying particle production as a function of 
the collision centrality. The determination of the latter is far from being trivial in $\pPb$ 
collisions because the intrinsic fluctuation of multiplicity in nucleon-nucleon collisions 
and the small number of nucleon-nucleon collisions imply only a broad correlation between
event ``activity'' and collision geometry. 
The experiments
developed different recipes~\cite{ALICEcentr,ChiaraOppedisano,ATLAScentr,Peripelitsa}. ALICE 
acknowledges the impossibility of an unbiased centrality determination and 
remarks this by defining $Q_{\rm pPb}$ in place of $\RpPb$. With the least bias estimator, based
on the measurement of and assumption of particle scaling, the charged
particle $Q_{\rm pPb}$ is compatible with unity for $\pt\gtrsim 8~\gev/c$~\cite{ALICEcentr,ChiaraOppedisano}. ATLAS does not modify
the determination of the estimated centrality and defines centrality bias correction factors to be applied to the yield of measured signals
on the basis of the correlation of the transverse energy deposited in the FCal calorimeter
and the average number of hard scatterings, for a given $N_{\rm coll}$~\cite{ATLAScentr}. With these correction 
factors the Z boson yield is found to scale with $\langle N_{\rm coll} \rangle$.

The rapidity distribution of Z bosons measured in $\pPb$ 
collisions by ATLAS shows an asymmetry between positive (Z boson going ``forward'' in p-going direction)
and negative (Z boson going ``backward'' in Pb-going direction) rapidities, with a small enhancement of the production
cross section at backward rapidity, slightly underestimated by theoretical models~\cite{ATLASplenary}. 

A complete diagnosis of $\Jpsi$ production in p--Pb collisions was carried out by all the four LHC experiments, investigating its dependence
on momentum, rapidity, and event multiplicity (aka centrality)~\cite{LHCbJPsipPb,ALICEJPsiQpPb,IgorLakomov,ATLASjpsipPbBrooks,CMSjpsipPbFilipovic}, setting stringent 
constraints to theoretical 
predictions. ALICE and CMS data indicate that cold nuclear matter effects are stronger in central 
than in peripheral p-Pb collisions and show that the suppression observed at forward rapidity (p-going direction) concerns
mainly the production of $\Jpsi$ with $\pt\lesssim 5~\gev/c$ in central collisions. No model is able to reproduce precisely
all the measured trends, though models using EPS09 nuclear PDF and a model based on coherent energy loss show generally a good
agreement with data within uncertainties.  ALICE observed also a suppression of $\psi(2{\rm S})$ production that increases 
with the collision centrality~\cite{IgorLakomov}, resembling what seen by PHENIX in d-Au collisions 
at RHIC~\cite{PHENIXpsi2sdAu}. The suppression in minimum bias data is described by a model including interactions 
of quarkonia with co-moving particles in the system. This ``final state'' effect could account also for the dependence 
of the suppression on the event multiplicity. A somehow puzzling enhancement of the production of high $\pt$ $\psi(2{\rm S})$  
was observed by ATLAS in peripheral collisions~\cite{ATLASjpsipPbBrooks}. More studies are needed to understand the origin of this observation.

Since the first tantalizing observation of a double-ridge structure in the angular correlation distribution of particles 
produced in p-Pb collision events with high multiplicity, resembling the elliptic-flow correlation typical of nucleus-nucleus
collisions, an effort was done by all the experiments for investigating its nature. ALICE presented the result of a new analysis 
in which muons reconstructed in the dedicated forward spectrometer are correlated with particle ``tracklets'' in the central 
barrel, thus separated by a large rapidity gap~\cite{ALICEmuonhcorrel}. Positive $\vtwo$ values were measured also for muons with transverse 
momentum larger than $2~\gev/c$, which predominantly come from decays of heavy-flavour hadrons. Slightly larger $\vtwo$ values were measured 
with muons at backward rapidity (Pb-going direction) than at forward rapidity. These results set important constraints to models in which 
a double-ridge structure in the angular correlation originates from effects related to the initial stage (e.g. from saturation of gluon 
nuclear parton distribution at low Bjorken x) or final stage (e.g. from a hydrodynamic evolution of the system) of the collision. A major result
for the investigation of the double-ridge in small systems was obtained by PHENIX with the observation of a large $v_{3}$ in 
${\rm He}^{3}$--$Au$ collisions, establishing a connection between the effect observed and the initial collision 
geometry~\cite{PHENIXheAuSickles}. Also PHENIX observed
larger $\vtwo$ values at backward rapidity, in the Au-going direction than at forward rapidity, in the direction of the lighter 
colliding ${\rm He}^{3}$ nucleus. 

The presence of the double-ridge in small systems, including high-multiplicity events in pp collisions, and 
the observation of a dependence of open charm, open beauty and quarkonia signals as a function of event multiplicity
in pp collisions, implies that particle production (including particles from hard-scattering processes) 
and their kinematic properties are connected to (soft?) processes characterizing the global event properties and structure. This 
``connectivity'' does not imply ``collectivity'', at least not in terms of a hydrodynamical evolution, and viceversa. Other processes
like multi-parton interactions with color reconnection can introduce long-range correlations. Connectivity could however be a seed for 
collectivity, though not necessarily the unique one: collective motion might develop in different ways 
in large and small systems (if any develops in the latter). A major goal in future years will be to clarify and understand
the origin of the ridge in small collision systems and what mechanisms relate the production rate of particles from 
hard-scatterings to event multiplicity. 







\end{document}